\documentclass[reviewcopy]{elsart}

\usepackage[latin1]{inputenc}
  \usepackage{newlfont}
  \usepackage{graphicx} 
  \usepackage{dcolumn}
  \usepackage{amssymb}
  \usepackage{amsfonts}
  \usepackage{amsmath}
  \usepackage[latin1]{inputenc}
  \usepackage{float}
\newcommand{\br}{\begin{eqnarray}}
\newcommand{\er}{\end{eqnarray}}
\newcommand{\nn}{\nonumber}
\newcommand{\bra}{\langle}
\newcommand{\ket}{\rangle}

\newcommand{\vt}{\vec}

\begin{document}

\begin{frontmatter}
\title{ Coherent state approach to the cross collisional effects in the population dynamics of a two-mode Bose-Einstein condensate
}

\author{Thiago F. Viscondi},
\author{K. Furuya}, 
\author{M. C. de Oliveira}
\address{Institute of Physics ``Gleb Wataghin'', University of Campinas, P. O. Box 6165, 13083-970, Campinas, SP, Brazil}

\begin{abstract}

We reanalyze the non-linear population dynamics of a Bose-Einstein Condensate (BEC) in a double well trap considering a semiclassical approach based on a time dependent variational principle applied to coherent states associated to SU(2) group. Employing a two-mode local approximation and hard sphere type interaction, we show in the Schwinger's pseudo-spin language the occurrence of a fixed point bifurcation  that originates a separatrix of motion on a sphere. This separatrix  corresponds to the borderline between two dynamical regimes  of Josephson oscillations and mesoscopic self-trapping. We also consider the effects of interaction between particles in different wells, known as cross collisions. 
 Such terms are usually neglected for traps sufficiently far apart, but recently it has been shown that they contribute to the effective tunneling constant with a factor growing linearly with the particle number. This effect changes considerably the effective tunneling of the system for sufficiently large number of trapped atoms, in perfect accord with experimental data. Finally, we identify analytically the transition parameter associated to the bifurcation in the generalized phase space of the model with cross-collision terms, and show how the dynamical regime depends on the initial conditions of the system and the collisional parameters values.
\end{abstract}


\end{frontmatter}

\section{Introduction}

Understanding of coherent quantum tunneling of matter waves is an
 important issue in physics. In particular, the dynamics of interacting fields
 can be extremely complex, giving birth to very counter intuitive phenomena like 
coherent non-spreading wave packets and nonlinear self-trapping. These phenomena 
have been clearly observed in the dynamics of atomic Bose-Einstein condensates 
(BEC) in double-well and periodic potentials \cite{shin,albiez,anker}, better 
than in superconducting Josephson junction arrays, thanks to the possibility to 
design well separated traps and due to the small dissipation in atom optical 
contexts. The wave tunneling in such system is analogous to the Josephson
effect \cite{josephson} and results in oscillatory exchange of the condensed 
neutral atoms between the adjacent traps, as has been suggested by several 
authors \cite{joscillations} . 
The nonlinear Josephson oscillations (JO) including the many-body interaction 
inside each trap for the atomic BEC in a double-well trap potential are 
discussed in \cite{milburn,smerzi,raghavan}, where the suppression of the 
oscillation, the so called {\sl self-trapping} phenomenon, happens when the 
initial population imbalance is above a threshold value. Based on the 
modeling of the Gross-Pitaevski equation using the mean-field  factorization 
and two-mode approximation, Milburn {\sl et al} \cite{milburn} presented an 
analytical solution for the dynamical transition to the mesoscopic self-trapping 
(MST) regime (as the number of particles exceeds a threshold value) in a 
particular initial condition where all the atoms are in one of the wells. 
More recently the possibility that the cross-collisional interaction become 
important for sufficiently large number of particles has been discussed in 
\cite{bruno}, hence altering the regime of transition from JO to MST. 

The purpose of this paper is first to present the complete mean field 
dynamics of the atomic BEC in two wells, complementing the work done in 
Milburn {\sl et al} \cite{milburn} by representing this integrable 
dynamics in terms of appropriate coherent states, which leads naturally
to a representation on a sphere \cite{anglin,kellman,trimborn}. We show 
all the relevant fixed points and its stability and discussing the physically 
permitted regimes compatible with the approximations taken to obtain the model. 
 This allows us to understand all the transitions 
between different dynamical regimes as a function of the collision parameters,
the tunneling coefficient and the number of atoms in the condensate. 
We also find the bifurcation condition for the dynamical fixed point of 
the mean field equations that determines the appearance or absence of the 
self-trapping regime in the system.
Secondly,  we generalize this treatment by including the collision
terms between atoms in different wells, owing to the overlap of the tails of 
the condensate wave functions in each trap, which is shown in \cite{bruno} 
that, depending on the number of atoms, this term is non-negligible. 
We also show the effect of the cross collisional terms on the dynamical 
transitions, thus determining the physical regime where it can be observed,
 a feature not realized in ref. \cite{kellman}.
  
The paper is organized as follows: in Section II we quickly show the two
mode approximation model for the Bose-Einstein condensate in a double well
\cite{milburn}, but including the cross-collisional terms \cite{bruno} and 
restate its Hamiltonian in terms of the Schwinger quasi-spin operators. 
Section III is devoted to the analysis of the dynamics of a classical 
analogue of this model. By studying the semi-classical Hamiltonian obtained
based on the time dependent variational principle (TDVP) applied to coherent 
states associated to SU(2) group, we localize all the fixed points of the 
model on the spherical phase space and its stability. By varying  one control 
parameter of the model we show a bifurcation of one of the fixed points on a 
sphere, which causes the appearance of a separatrix of motion, dividing the 
phase space in two regions of different dynamical regimes corresponding to 
the Josephson oscillations and mesoscopic self-trapped motions of the condensate. 
In section IV we show the connection of the classical mean field bifurcation
with the quantum dynamical behavior, illustrated by Husimi distributions,
 and the spectra underlying such behavior.
Also, we show how  such a dynamical structure 
on the sphere is modified when the cross-collisional terms are included. 
Finally in section V we present our conclusions.

\section{The two-mode local approximation (TMLA)}

\begin{figure}[h]
\begin{center}
\centering 
\includegraphics[scale=0.5]{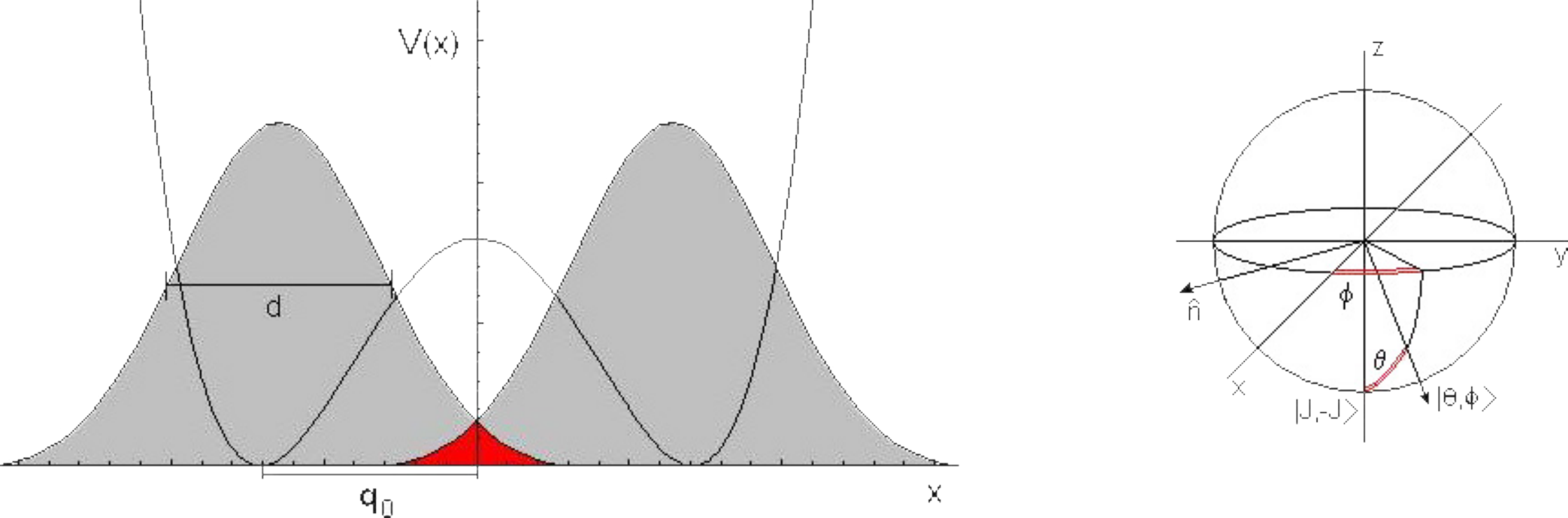}
\end{center}
\caption{Left side: scheme of the double well potential with localized wave functions $u_{\pm}(\vec{r})$. Right side: Bloch sphere with the definitions of the angles parametrizing the atomic coherent states. Notice that the angle $\theta$ is different from the one of usual spherical coordinates.}
\label{fig1}
\end{figure}

Considering a trapping potential $V(\vec{r})$ with two equivalent global minima
at $\vec{r}_{\pm}= \pm q_0 \hat{x}$, for simplicity, we assume the potential to 
be harmonic with frequency $\omega$ in both axis $y$ and $z$ and, without 
loss of generality, having null value at the minima. In this way we obtain the 
following single particle Hamiltonian with mass $m$:
\br
H=\frac{\vt{p}^{2}}{2m}+V(\vt{r})=\frac{\vt{p}^{2}}{2m}
+\frac{b}{q_{0}^{2}}(x^{2}-q_{0}^{2})^{2}+\frac{m\omega^{2}}{2}(y^{2}+z^{2}),
\label{Hsingle}
\er
where $b$ is an adjustable parameter associated to the height of the potential between the two wells, which we choose as $b=\frac{m \omega^2}{8}$ in such a way that the harmonic approximation $V_{\pm}^{(2)}= \frac{m\omega^{2}}{2}\left[(x\mp q_{0})^{2}+y^{2}+z^{2}\right]$ of the potential over each minimum becomes isotropic. If the distance $2q_0$ between the minima is large enough we can suppose that the harmonic approximation around each minima acts as independent trapping potential to the more populated low energy states of the BEC. Thus, we can suppose that the two practically degenerated ground states of the global potential are the symmetric and anti-symmetric combinations of the approximated harmonic ground states, given by simple Gaussians $\langle \vec{r}|u_{\pm}\rangle=\frac{1}{\pi^{\frac{3}{4}}d^{\frac{3}{2}}}e^{-\frac{[(x\mp q_{0})^{2}+y^{2}+z^{2}]}{2d^{2}}}$ with width $d=\sqrt{\frac{\hbar}{m\omega}}$, as in Fig.(\ref{fig1}).

Now consider a system of $N$ interacting bosons, where the Hamiltonian in a second quantized form \cite{negele} can be written as 
\br
\hat{H}& = &\int
d^{3}r\hat{\psi}^{\dag}(\vt{r})H(\vt{r})\hat{\psi}(\vt{r})
\nn \\ 
& &+\frac{1}{2}\int d^{3}r d^{3}r'
\hat{\psi}^{\dag}(\vt{r})\hat{\psi}^{\dag}(\vt{r}')V'(\vt{r},\vt{r}')\hat{\psi}(\vt{r}')\hat{\psi}(\vt{r}).
\label{hamil1}
\er
Then, supposing that the single level energy spectra is not considerably affected by the interaction, a TMLA is done: $\hat{\psi}(\vt{r})\approx \bra \vt{r}|u_{+}\ket d_{+}+\bra \vt{r} |u_{-}\ket d_{-}$, where $d_{\pm}$ are boson annihilation operators on the states $|u_{\pm}\rangle$. We also approximate the interaction potential by a hard sphere type  potential for low energy particles : $V^{'} (\vec{r},\vec{r}^{'})=V_o \delta(\vt{r}-\vt{r}')= \frac{4\pi\hbar^{2}a}{m} \delta(\vt{r}-\vt{r}')$, where $a$ is the $s$-wave scattering length \cite{pethick}. By inserting the above approximations in Eq.(\ref{hamil1}) and introducing the overlap $\epsilon \equiv \bra u_{+}|u_{-}\ket=e^{-\frac{q_{0}^{2}}{d^{2}}} $, which is small ($\epsilon \ll 1$), we define the tunneling parameter $\Omega = 2\langle u_{-}|H|u_{+}\rangle$ and the collision parameters $\kappa \equiv \frac{V_o}{2}\int d^3r u_i^4 (\vec{r})$, $\eta \equiv \frac{V_o}{2}\int d^3r u_i^2 (\vec{r})u_j^2 (\vec{r}) = \kappa \epsilon^2$ and $\Lambda \equiv \frac{V_o}{2}\int d^3r u_i^3 (\vec{r})u_j (\vec{r}) = \kappa \epsilon^{3/2}$ for $i \ne j$ ($i,j=+,-$). Discarding constant terms we get the two-mode Hamiltonian which describes the BEC in a double well. However, to explore the natural group structure of the model (because of the fixed number of particles $N$) it is more convenient to adopt the pseudo-spin operators approach introduced by Schwinger \cite{schwinger} by defining the following operators:
\br
J_{x}&\equiv& \frac{d_{-}^{\dag}d_{-}-d_{+}^{\dag}d_{+}}{2},\quad
J_{y}\equiv i\frac{d_{-}^{\dag}d_{+}-d_{+}^{\dag}d_{-}}{2}\quad
\mbox{and}\quad
 J_{z}\equiv \frac{d_{+}^{\dag}d_{-}+d_{-}^{\dag}d_{+}}{2}, 
\label{Jschwinger}
\er
with $J=N/2$. 
{ In the many-body theory $J_x$ can be interpreted as proportional to the position of the condensate in the $x$-axis;  $J_y$ as the linear momentum of the condensate along the same axis and $J_z$ as the population imbalance between the symmetric and antisymmetric energy eigenstates.} The Hamiltonian (\ref{hamil1}) in terms of these operators assumes the following form:
\br
\hat{H} = 2\left[2\Lambda(N-1)+\frac{\Omega}{2}\right]J_{z}+2(\kappa-\eta)J_{x}^{2}+4\eta
J_{z}^2.
\label{hamil2}
\er
This is a Lipkin-Meshkov-Glick (LMG) type Hamiltonian, which has been widely 
discussed in the literature \cite{lipkin,vidal}. 
Here, $\kappa$ is the self-collision parameter of the condensate and, as can be seen from its definition, it is much larger than the so called cross collision terms $\eta,\, \Lambda$ (non-independent),
 since $\epsilon \ll 1$ \cite{milburn}. However, we cannot neglect these lower order parameters, as discussed in \cite{bruno}, due to its presence in the tunneling term of the Hamiltonian though the effective tunneling parameter $\Omega^{'}\equiv 2[2\Lambda(N-1)+ \Omega/2]$, where $\Lambda$ (although small) is multiplied by an extra factor proportional to $N$, which in a typical experiment can range from $10^2$ to $10^{10}$.
 Indeed this dependence of the tunneling rate on the number of atoms has shown to be quite relevant for the indirect relative phase inference of one of the two-mode condensates through the atomic homodyne detection as proposed in Ref. \cite{bruno2}.
\section{A Classical Analogue}

In this section we obtain the semiclassical Hamiltonian of the model, based on a TDVP \cite{saraceno} via the $Q$-representation of the Hamiltonian (\ref{hamil2}) $\langle \theta,\phi| \hat{H}| \theta,\phi\rangle $ in the so called atomic ($SU(2)$) coherent states \cite{arecchi}. This is given by:
\begin{equation}
 |\theta,\phi\ket \equiv
e^{-i\theta(J_{x}\sin\phi-J_{y}\cos\phi)}|J,-J\ket=\sum_{M=-J}^{J}\sqrt{{2J}\choose
{M+J}}\frac{\tau^{M+J}}{(1+|\tau|^{2})^{J}}|J,M\ket,
\label{cohst}
\end{equation}
where in the last equality $\tau \equiv e^{-i\phi} \tan{}\frac{\theta}{2}$ with
$\theta,\phi$ as defined in Fig.(\ref{fig1}).
Using the above expansion in the angular momentum basis, we obtain
\br
\langle \theta,\phi| \hat{H}| \theta,\phi\rangle \equiv {\mathcal H}(\tau,\tau^*)&=& -J\Omega'\frac{1-\tau^{*}\tau}{1+\tau^{*}\tau}
+(\kappa-\eta)\frac{J(2J-1)}{(1+\tau^{*}\tau)^{2}}(\tau^{*}+\tau)^{2}
\nn \\& & 
-8\eta\frac{J(2J-1)\tau^{*}\tau}{(1+\tau^{*}\tau)^{2}}.
\label{hamil3}
\er
 Then, we get the following canonical Hamilton's equation of motion for the generalized variables $(q,p)$ (with $\tau=\frac{q+ip}{\sqrt{4J-q^2-p^2}} $) in the phase space 
\br
\dot{q}&=&\Omega'p-(\kappa-\eta)\frac{(2J-1)}{2J}q^{2}p+
\eta\frac{(2J-1)}{J}p(2q^{2}+2p^{2}-4J),\nn
 \\
\dot{p}&=&-\Omega'q-(\kappa-\eta)\frac{(2J-1)}{2J}q(4J-2q^{2}-p^{2})\nn\\
&&-\eta\frac{(2J-1)}{J}q(2q^{2}+2p^{2}-4J) .
\label{eqmotion}
\er
By imposing $\dot{q}=\dot{p}=0$ in (\ref{eqmotion}), and transforming the canonical variables to the Bloch sphere angles $\theta,\phi$, we obtain the following four sets of fixed points:
\br
&&\mbox{(a)}\,\, \theta  =  0,\,\, \phi\,\, \mbox{undetermined  (south pole)} ;\nn\\
&&\mbox{(b)}\,\,\theta  =  2 \arctan{\left(\sqrt{\frac{R^2(k-n)+\frac{\Omega^{'}}{2} }{R^2(k-n)-\frac{\Omega^{'}}{2} }}\right)},\,\, \phi= \left\{ \begin{array}{c} 0, \,\, \mbox{for $q>0$} \\ 
\pi,\,\,  \mbox{for $q<0$}\end{array} \right.;\nn\\
\label{fixpoint} 
&&\mbox{(c)} \,\, \theta  =  2 \arctan{\left(\sqrt{\frac{R^2 n+\frac{\Omega^{'}}{2} }{R^2 n -\frac{\Omega^{'}}{2} }}\right)},\,\, \phi= \left\{ \begin{array}{c} \pi/2, \,\, \mbox{for $p<0$} \\ 
-\pi/2, \,\, \mbox{for $p>0$} \end{array} \right.;\nn\\  
&&\mbox{(d)}\,\, \theta  =  \pi, \,\, \phi =  \left\{ \begin{array}{c}
 + \arctan{\left(\sqrt{\frac{R^2(k-n)-\frac{\Omega^{'}}{2} }{R^2n+\frac{\Omega^{'}}{2} }}\right)},\,\, \mbox{for $qp <0$} \\
 - \arctan{\left(\sqrt{\frac{R^2(k-n)-\frac{\Omega^{'}}{2} }{R^2n+\frac{\Omega^{'}}{2} }}\right)},\,\, \mbox{for $qp > 0$} \end{array}  \right.; \nn
\er
where we defined $k\equiv (\kappa - \eta)\frac{2J-1}{4J}$ and $n\equiv \eta \frac{2J-1}{2J}$ and $R\equiv 2 \sqrt{J}$. 

Now, analyzing the stability \cite{alfredo} of the above points  and considering the  parameter regimes of the system compatible with the approximations discussed in Sec. II, i. e.,
\begin{equation}
\Omega,\kappa \gg \Lambda, \eta \,\,(\Lambda>\eta)
\label{restrict}
\end{equation}
we show that:
\textit{(i)} The trivial fixed point \ref{fixpoint}(a) at the south pole ($\theta=0$), is 
{\em stable} in all the domain of parameters allowed by condition (\ref{restrict}).

\textit{(ii)} Fixed points \ref{fixpoint}(b) are only defined if $R^2(k-n)\ge \frac{\Omega^{'}}{2}$ and are {\em stable} except along the {\em critical line} $R^2(k-n)= \frac{\Omega^{'}}{2}$.

\textit{(iii)} Fixed points \ref{fixpoint}(c) exists only if $R^2n \geq \frac{\Omega^{'}}{2}$, 
 and are {\em stable} except if $p=0$ (reducing to the trivial solution). However,
this is physically non-achievable for the above conditions (\ref{restrict}).

\textit{(iv)} All the fixed points \ref{fixpoint}(d) correspond to a single point on the Bloch sphere $\theta=\pi$ (the north pole of the sphere) and are {\em unstable} for  $R^2(k-n)\ge \frac{\Omega^{'}}{2}$. Thus, when the two fixed points \ref{fixpoint}(b) become stable and depart from the north pole, the fixed point \ref{fixpoint}(d) becomes unstable, thus characterizing a bifurcation. 

The {\em bifurcation condition} obtained above, $R^2(k-n)= \frac{\Omega^{'}}{2}$, as expressed in terms of the parameters of the Hamiltonian is given by:
\begin{gather}
 (\kappa - 3 \eta)(N-1) = 2 \Lambda(N-1)+ \frac{\Omega}{2}. \label{stabcond}
\end{gather} 
%

\section{Dynamical Transition between JO and MST regimes}

\begin{figure}[ht]
\centering
\includegraphics[scale=0.165]{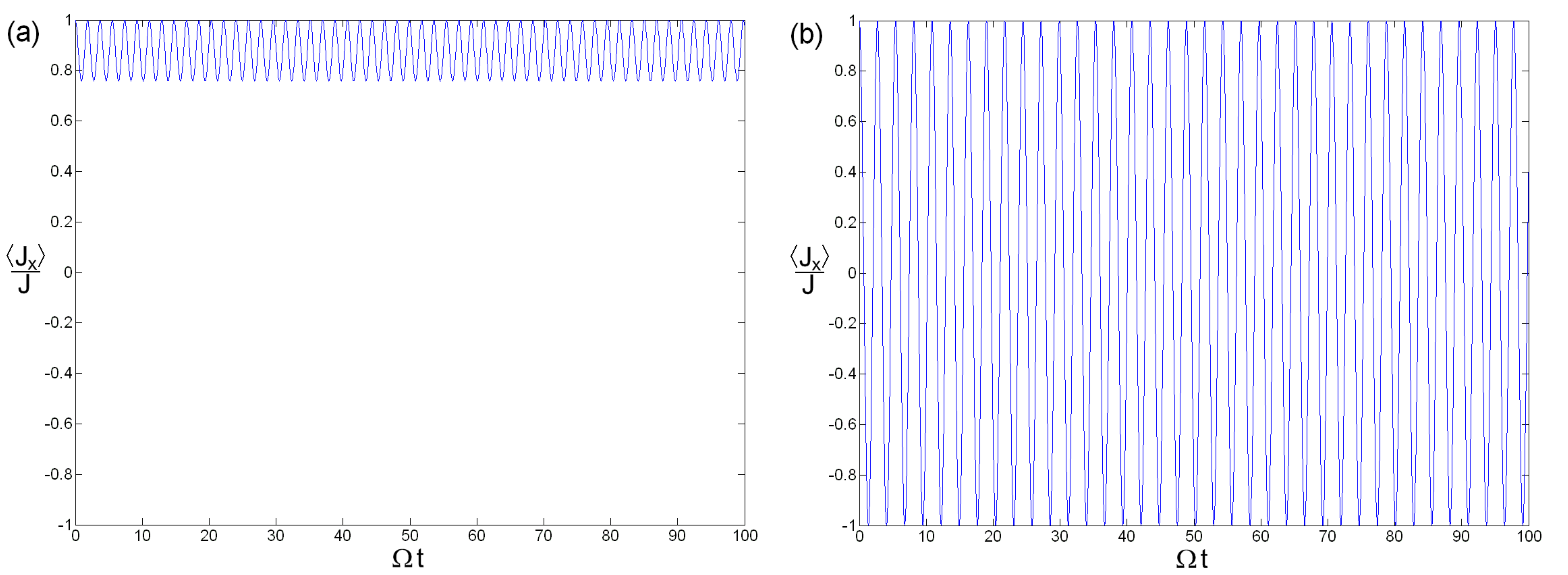}
\caption{ Semiclassical time evolution of $\langle \frac{J_x}{J} \rangle( \theta,\phi )$ in the MST and JO regimes for initial condition $\theta= \frac{\pi}{2}$ and $\phi=0$, corresponding to all atoms initially in one of the wells. The parameter values are chosen $N=100$, $\kappa= \frac{2 \Omega}{N}$ and the plots are in terms of adimensional parameter $\Omega t$, with cross collision parameter (a) $\eta= \kappa/100$; (b) $\eta= \kappa/10$.
}
\label{fig2}
\end{figure}

Now we discuss the quantum dynamics associated to the mean field classical analogue of the previous section. From the stability condition (\ref{stabcond}) obtained above, in the absence of cross-collision terms the {\em bifurcation condition} would give
\begin{gather}
 \kappa (N-1)= \frac{\Omega}{2}.
\end{gather} 
This corresponds to the transition between different dynamical regimes of the model described in \cite{milburn,fu} \footnote{ Note that the transition condition is {\em not} the same as in \cite{milburn}, since there it is chosen to fit a particular set of initial conditions, whereas here we choose to consider the exact point of transition of the system.}. 
 In the JO regime the position of the condensate, given by $\langle J_x \rangle$, oscillates in time around zero, indicating no preferential tunneling to any of the wells. In the MST regime, $\langle J_x \rangle$ oscillates around a non-zero value, showing that part of the condensate is mesoscopically trapped in one of the wells, as shown in Fig.\ref{fig2}.

\begin{figure}[ht]
\centering
\includegraphics[scale=0.165]{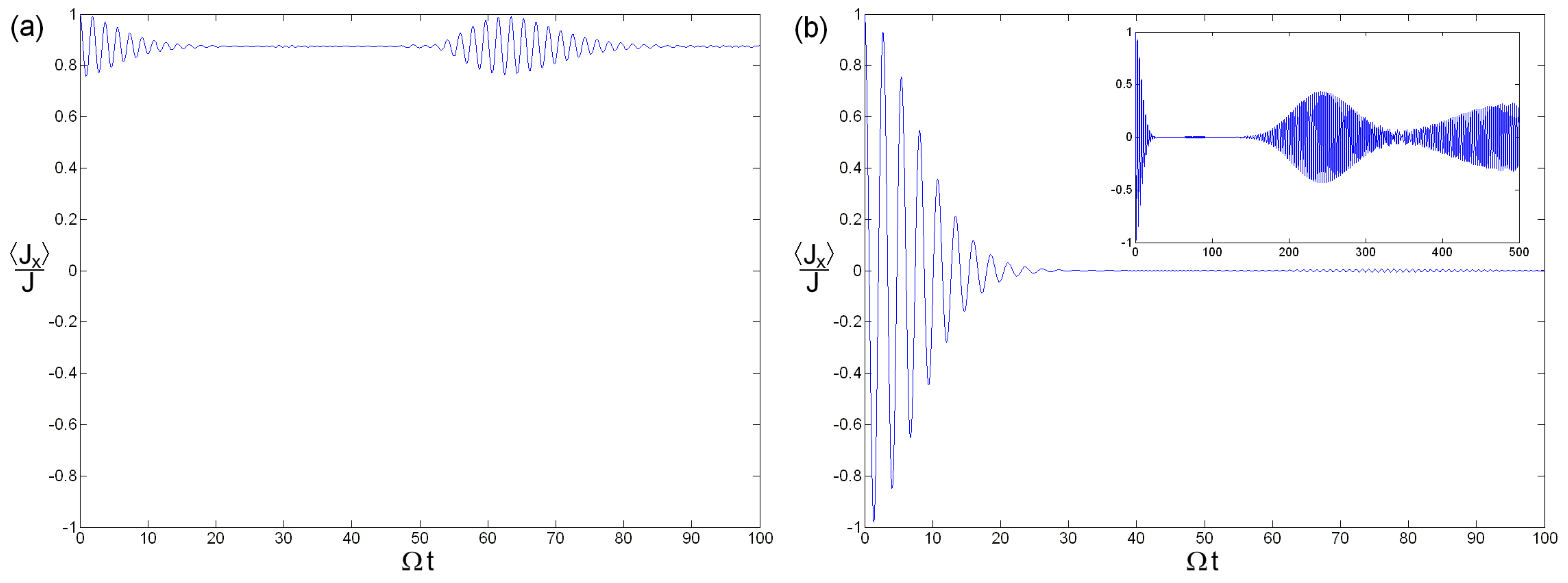}
\caption{ Quantum mechanical time evolution of $\langle \theta,\phi| \frac{J_x}{J}| \theta,\phi\rangle $ in the MST and JO regimes for the same initial condition and parameters of Fig.(\ref{fig2}) as a function of  $\Omega t$. (a) $\eta= \kappa/100$; (b) $\eta= \kappa/10$.
} 
\label{fig3}
\end{figure}

In Fig.(\ref{fig2}) (a) and (b) we show the semiclassical time evolution of  $\langle  \frac{J_x}{J}\rangle (t) $ for the same parameter values corresponding to the quantum mechanical results in Fig.(\ref{fig3})(a) and (b) respectively. It is clear that for $\eta= \kappa/100$ the initial state is related to a MST orbit whereas for  $\eta= \kappa/10$ it is related to a JO orbit. The semiclassical evolution is very similar to the quantum one, except for a modulation in the quantum case which produces a sequence of collapses and revivals (for $\eta= \kappa/10$, see the inset of Fig.(\ref{fig3})(b)).

\begin{figure}[ht]
\centering
\includegraphics[scale=0.2]{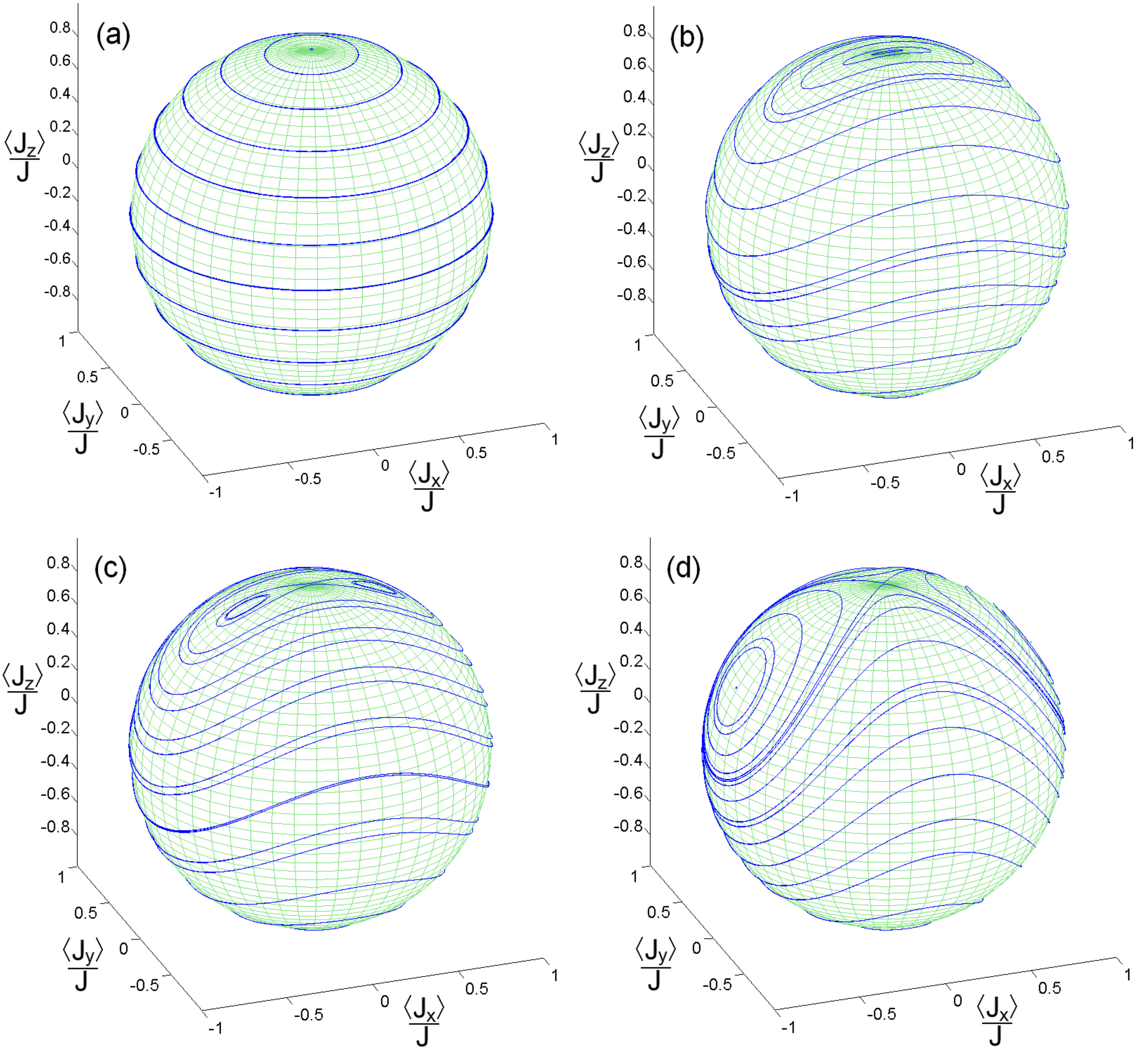}
\caption{ Trajectories on the Bloch sphere for several values of self-collision, and {\sl without} cross collision term ($\eta=0$). Fixing $N=100$ and $\Omega=1$, the self-collision parameters of each plot is given by: (a) $\kappa= 0$, (b) $\kappa= \frac{1.0}{2N}$,  (c) $\kappa= \frac{1.1}{2N}$ and (d) $\kappa= \frac{1.0}{N}$.
}
\label{fig4}
\end{figure}

We can better visualize  the orbits of the system in the unit sphere whose surface is parametrized by the polar angles $\theta$ and $\phi$, the so called Bloch sphere. Considering $\Omega > 0$ and $\kappa > 0$, we have a fixed point at $\theta = \pi$ (north pole) which is stable whenever $(\kappa-3 \eta)(N-1)< 2\Lambda(N-1)+ \frac{\Omega}{2}$. When this inequality is reversed, the north pole becomes unstable and two new fixed points appear at
 $\theta = 2 \arctan{\left(\sqrt{
\frac{R^2(k-n)+\frac{\Omega^{'}}{2} }{R^2(k-n)-\frac{\Omega^{'}}{2} }
}\right)}$ 
and $\phi=0,\pi$; characterizing a {\sl pitchfork type bifurcation}, as can be seen in Fig.(\ref{fig4}).

 Notice that in the case $\eta=0$, for increasing values of $\kappa$, before the transition point $\kappa_c = \frac{\Omega}{2(N-1)}$ there are solely orbits associated to the JO of the condensate. When the self-collision reaches the critical value, the bifurcation happens and orbits of MST orbits appears, occupying an increasing area in the phase space of the system as $\kappa$ increases.  As already noticed in previous publications for particular initial conditions \cite{milburn}, the increase of collision rates between bosons in the same site are responsible for the phenomenon of suppression of tunneling. 

The $Q$-representation of a Hamiltonian gives the exact classical limit of a system in the $N \rightarrow \infty$ limit \cite{zhang}, thus the {\sl bifurcation condition} on the parameters  $(\kappa-3 \eta)(N-1)= 2\Lambda(N-1)+ \frac{\Omega}{2}$
represents in this limit the exact transition condition between the two dynamical regimes of the model.

\begin{figure}[ht]
\centering
\includegraphics[scale=0.16]{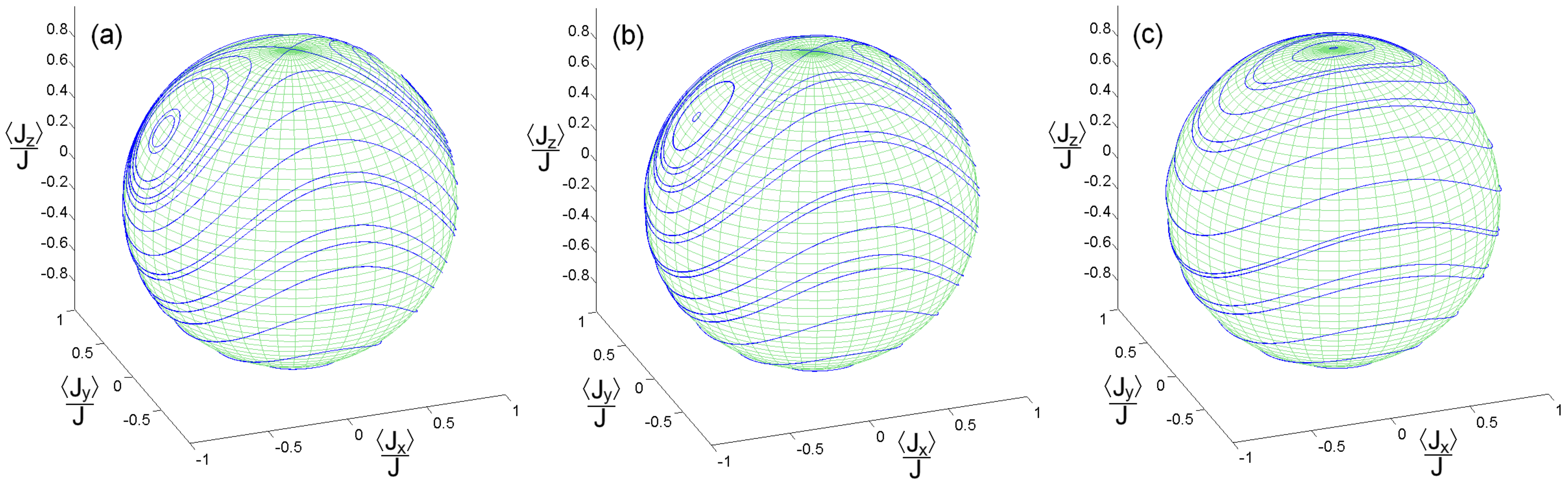}
\caption{Trajectories on the Bloch sphere, considering cross collision terms with increasing values of $\eta$ and fixed parameter values $N=100$, $\Omega=1$ and $\kappa=\frac{1}{N}$. (a) $\eta=\frac{\kappa}{100}$, (b) $\eta=\frac{\kappa}{40}$, (c) $\eta=\frac{\kappa}{10}$.
}
\label{fig5}
\end{figure}

Now, we consider the effect of the cross collisional terms on the population dynamics, and we find a non-negligible change in the effective tunneling even for a number of particles not so large ($N=100$) as shown in Fig.(\ref{fig5}).

 In opposition to the self-collision term, the increase of $\eta$ suppresses the presence of the MST orbits from the phase space. Even for values of $\eta$ much smaller than $\kappa$ we observe a significant change in the dynamical behavior of the system; namely, for solely $100$ particles it is already noticeable the modification in the effective tunneling of the system. This fact goes in the same direction of the experimentally observed situation\footnote{ This does not exclude other possible explanations for these effects such as the one presented in \cite{andrea}.} \cite{albiez}. Notice that in this work we always keep the cross-collision rates considerably lower than the self-collision rates, respecting the hypothesis used to obtain the model, and we select only the {\sl physically allowed bifurcations} from those listed in \cite{kellman} obtained from Eq.(\ref{eqmotion}).

\begin{figure}[ht]
\centering
\includegraphics[scale=0.15]{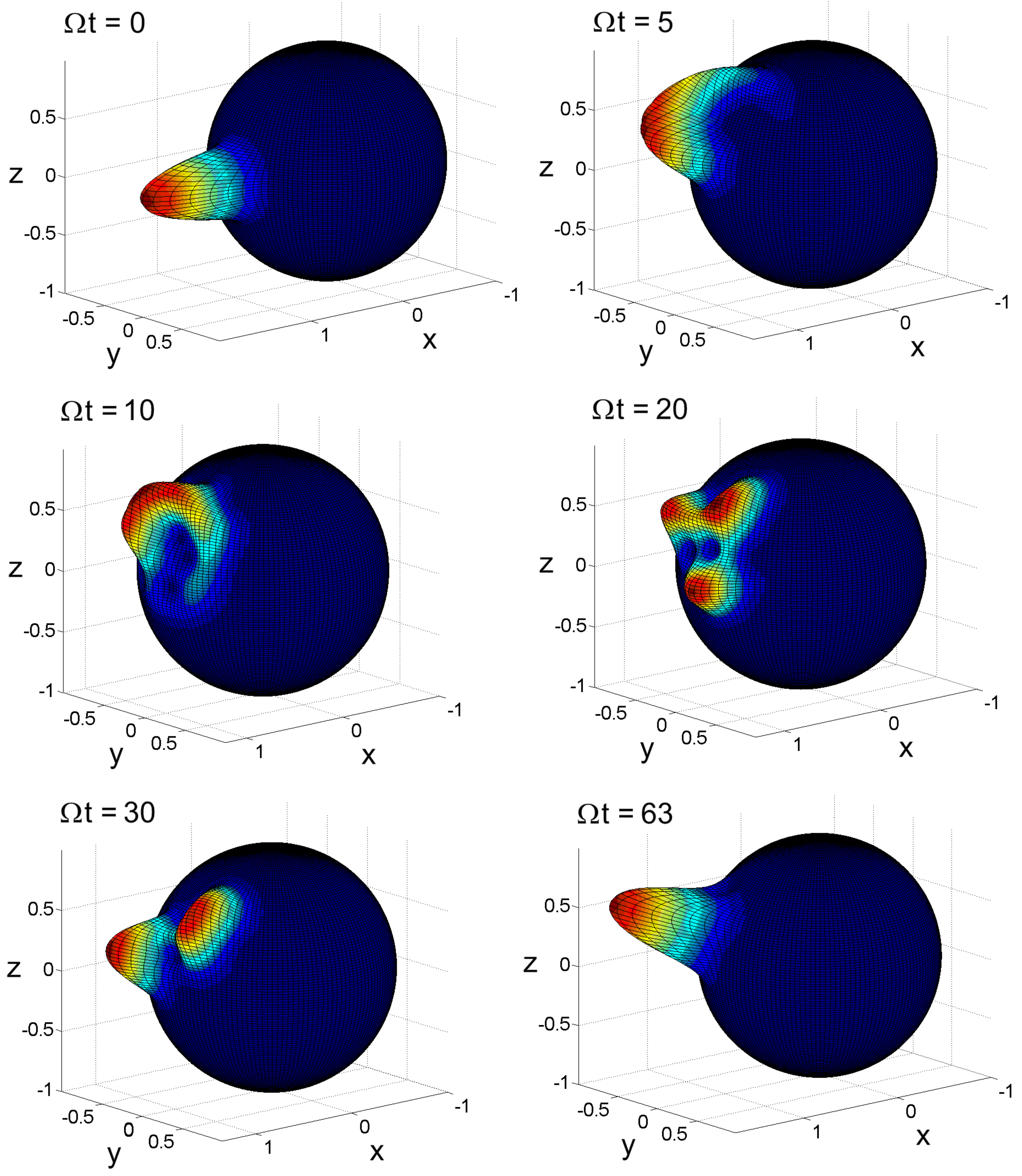}
\caption{Plot of a time evolved Husimi functions for MST initial condition
with parameter values $N=100$, $\eta=\kappa/100$ and $\kappa = 1/N$
for $\Omega t = 0,5,10,20,30,63$.
}
\label{fig6}
\end{figure}

\begin{figure}[ht]
\centering
\includegraphics[scale=0.165]{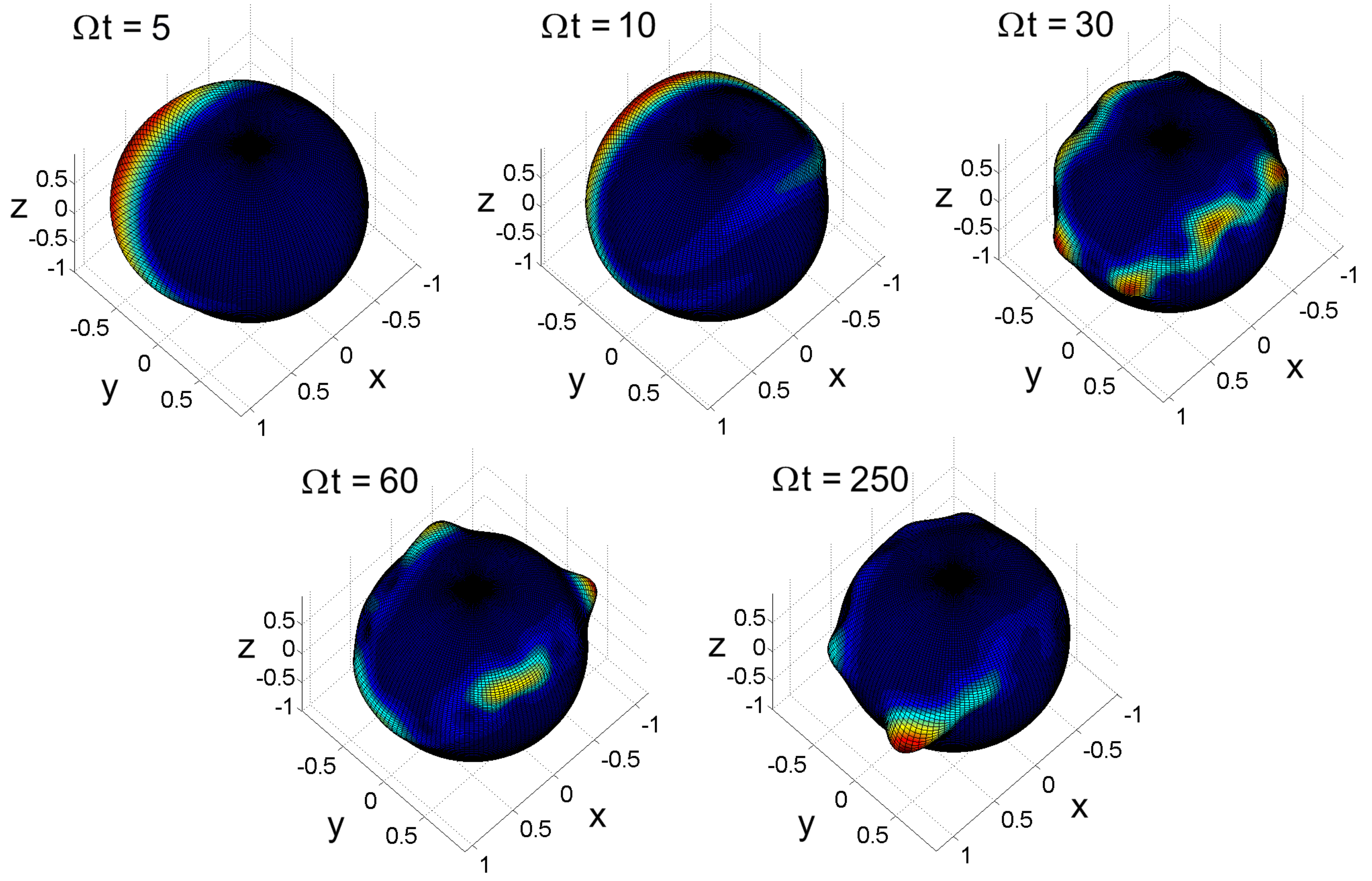}
\caption{Plot of a time evolved Husimi functions for JO initial condition
with parameter values $N=100$, $\eta=\kappa/10$ and $\kappa = 1/N $
for $\Omega t = 5,10,30,60,250$.
}
\label{fig7}
\end{figure}

  In order to show the consistency of the quantum dynamics with both 
dynamical regimes, we show in Figs.(\ref{fig6}) and (\ref{fig7}) a sequence 
of plots of Husimi functions on the sphere for time evolved states with the 
same initial coherent state, but for parameter values corresponding to MST and 
JO regimes, respectively.

  Although the initial state for Figs (\ref{fig6}) and (\ref{fig7}) is the 
same coherent state on the sphere centered at the position where the $x$-axis 
crosses the unit sphere (represented in Fig. (\ref{fig6}) for $\Omega t=0$) the
 time evolved states are completely different due to their respective dynamical
 regimes. Fig.(\ref{fig6}) corresponds to the MST regime as shown in 
Fig.(\ref{fig3}a) with the classical trajectories in phase space corresponding 
to the one shown in Fig(\ref{fig5}a).
Thus, the time evolution shows first a spreading of the 
Q-function along the classical trajectory ($\Omega t=5,10$) in the trapped
region of the phase space, and then shows effects of self-interferences 
($\Omega t=20,30$) forming mesoscopic superpositions with three and two well 
defined peaks, and finally re-constructs the one-peak quasi-coherent state at 
$\Omega t = 63$.
Notice that $\Omega t=30$ corresponds to the time at
which a small recovering of oscillations (revival) happens in the middle of the
collapse  and $\Omega t=63$ to the revival time in Fig.(\ref{fig3}a).  This 
behavior is a clear demonstration of the typical collapse and revival of the 
condensate state phase.

In Fig.(\ref{fig7}) on the other hand, it is shown the evolution corresponding
to the JO regime as shown in Fig.(\ref{fig3}b) and the classical trajectories
in phase space corresponding to the one shown in Fig(\ref{fig5}c). The initial distribution at $\Omega t=0$ is the same as in fig. (\ref{fig6}). One can see 
the spreading (collapse) of the distribution surrounding all the sphere 
characterizing the JO ($\Omega t= 5,10$), and then the self-interferences 
also happening but in a less coherent way ($\Omega t=30,60$). 
Finally at $\Omega t =250$ the best re coherence of this case is shown, but 
clearly in a much less coherent reconstruction, as expected due to the much 
longer classical trajectory to be traversed, and this is consistent with the 
less perfect revival in the inset of Fig.(\ref{fig3}b). 

Another signal of the dynamical phase transitions (JO $\rightarrow$ MST) can be seen in the energy spectra of the Hamiltonian (\ref{hamil2}). 
\begin{figure}[ht]
\centering
\includegraphics[scale=0.17]{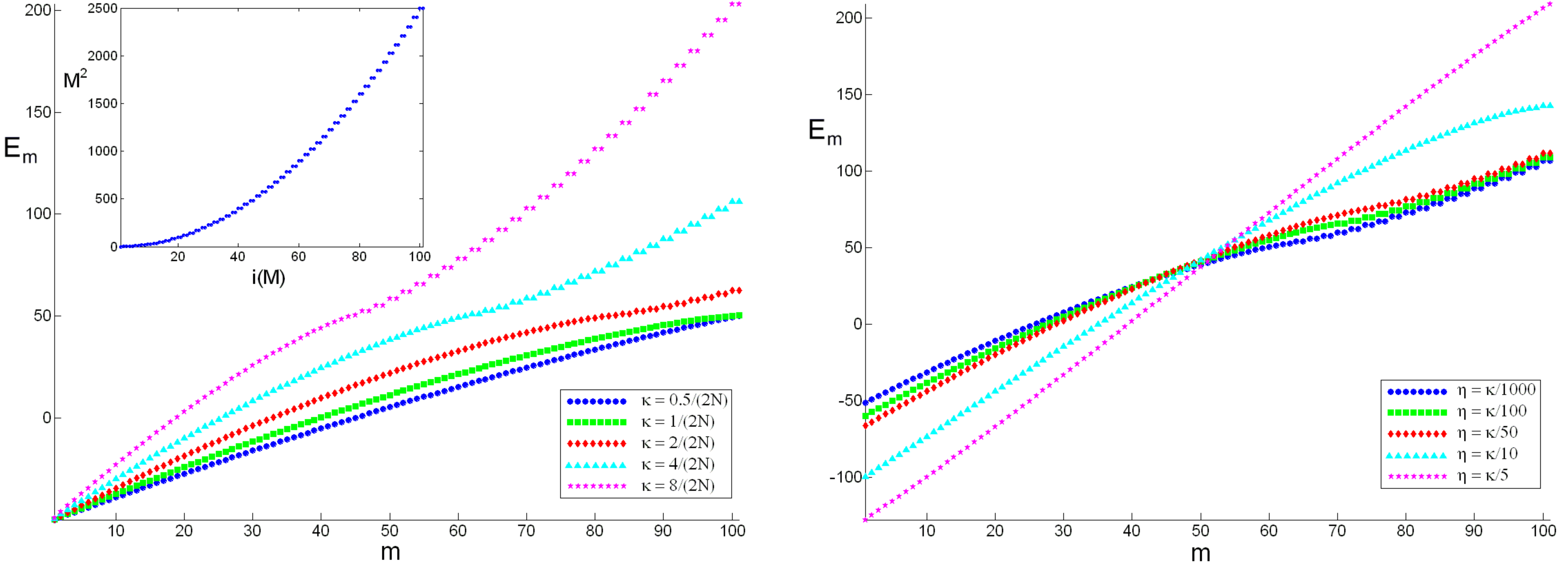}
\caption{Energy spectra $E_n$ as a function of the label $n$
for fixed parameter values $N=100$ and $\Omega=1$. (a) without cross-collision terms ($\eta=0$), and several values of self-collision parameter $\kappa$. (b) with cross-collision terms, fixing $\kappa=\frac{2}{N}$ and several values of $\eta$.
}
\label{fig8}
\end{figure}
In the spectra shown in Fig.(\ref{fig8})(a) we observe that for self-collision parameter above the critical value there is the appearance of an inflection point (\cite{inflection}) and a doublet structure. The term responsible for the doublet structure is the $J_x^2$ term in the Hamiltonian (\ref{hamil2}) (as shown in the inset) representing the self-trapped states when it is dominant (namely, eigenstates of well defined difference of the number of particles between the two wells). In Fig.(\ref{fig8})(b) we show the spectra with fixed $\kappa=\frac{2}{N}$ and several values of cross collision parameter $\eta$. As we increase $\eta$ (but keeping $\kappa$ significantly larger than $\eta$), the inflection point and the doublet structure are clearly suppressed. 
\section{Conclusion}

To conclude, we confirm first that the effect of cross-collision in the two mode approximation model for the dynamics of the Bose-Einstein condensate in a double well trapping potential should not be neglected in all circumstances \cite{bruno}. This is so because the intensity of this effect on the effective tunneling of the system increases linearly with the number of trapped particles, which in a typical experiment can easily reach a mesoscopic amount of atoms. Such effects has been shown in the simulations presented in Figures (\ref{fig4},\ref{fig7},\ref{fig8}) and these are qualitatively in accordance with existing experimental results \cite{albiez}. 

We also showed that the transition of the dynamical regime of the system is associated with a bifurcation in the generalized phase space in a semiclassical approach of the problem, in such a way that the dynamics of the condensate -- Josephson oscillation or macroscopic self-trapping -- is extremely dependent on the initial conditions chosen for the populations of the wells and its phase relations. Our semiclassical approach using PVDT and SU(2) coherent state guarantees both the correct phase space topology \footnote{The use of correct topology is essential to obtain the correct number of fixed points and to avoid a wrong interpretation of some limiting phase space orbits in flat spaces as nonexistent separatrices of motion.} and the particle number conservation. Also, taking advantage of the SU(2) group structure of the model leads naturally to a phase space motion on a sphere, facilitating the analysis and visualization of the dynamical structure associated to the nonlinear quantum dynamics of the double well condensate, namely JO or MST.    

 Finally, we show a situation where the cross-collision terms can even suppress the bifurcation and thus the appearance of the self-trapping regime in the Bose-Einstein condensate in a double well. 


\section*{Acknowledgments}
We thank J. Vidal for bringing aspects of the LMG model to our attention, and acknowledge support from Fun\-da\-\c{c}\~ao de Amparo \`a Pesquisa do 
Estado de S\~ao Paulo (FA\-PESP) (Proc. No. 2006/05142-4) and Conselho 
Nacional de Desenvolvimento Cient\'{\i}fico e Tecnol\'ogico (CNPq).

\end{document}